\documentclass[10pt]{iopart}
\usepackage{graphicx,iopams}

\begin{document}
\jl{12}
\submitted

\paper[Dynamic splitting of pencil beams for heterogeneity corrections]{Dynamic splitting of Gaussian pencil beams in heterogeneity-correction algorithms for radiotherapy with heavy charged particles}
 
\author{Nobuyuki~Kanematsu$^{1,2}$, Masataka~Komori$^{1,3}$, Shunsuke~Yonai$^1$ and Azusa~Ishizaki$^{1,2}$}

\address{$^1$ Department of Accelerator and Medical Physics, Research Center for Charged Particle Therapy, National Institute of Radiological Sciences, 4-9-1 Anagawa, Inage-ku, Chiba 263-8555, Japan}

\address{$^2$ Department of Quantum Science and Energy Engineering, School of Engineering, Tohoku University, 6-6 Aramaki Aza Aoba, Aoba-ku, Sendai 980-8579, Japan}

\address{$^3$ Department of Radiological Technology, School of Health Sciences, Nagoya University, 1-1-20 Daiko Minami, Higashi-ku, Nagoya 461-8673, Japan}

\eads{nkanemat@nirs.go.jp}

\begin{abstract}
The pencil-beam model is valid only when elementary Gaussian beams are small enough with respect to lateral heterogeneity of a medium, which is not always the case in heavy charged particle radiotherapy.
This work addresses a solution for this problem by applying our discovery of self-similar nature of Gaussian distributions.
In this method, Gaussian beams split into narrower and deflecting daughter beams when their size has exceeded the lateral heterogeneity limit.
They will be automatically arranged with modulated areal density for accurate and efficient dose calculations.
The effectiveness was assessed in an carbon-ion beam experiment in presence of steep range compensation, where the splitting calculation reproduced the detour effect of imperfect compensation amounting up to about 10\% or as large as the lateral particle disequilibrium effect.
The efficiency was analyzed in calculations for carbon-ion and proton radiations with a heterogeneous phantom model, where the splitting calculations took about a minute and were factor of 5 slower than the non-splitting ones.
The beam-splitting method is reasonably accurate, efficient, and general so that it can be potentially used in various pencil-beam algorithms.
\end{abstract}


\section{Introduction}

In treatment planning of radiotherapy with protons and heavier ions, the pencil-beam (PB) algorithm is commonly used (Hong \etal 1996, Kanematsu \etal 1998, 2006, Schaffner \etal 1999, Kr\"amer \etal 2000), where a radiation field is approximately decomposed into two-dimensionally arranged Gaussian beams that receive energy loss and multiple scattering in matter.
In the presence of heterogeneity, these beams grow differently to reproduce realistic fluctuation in the superposed dose distribution.

Comparisons with measurements and Monte Carlo (MC) simulations, however, revealed difficulty of the PB algorithm at places with severe lateral heterogeneity such as steep areas of a range compensator and lateral interfaces among air, tissue, and bone in a patient body (Goitein 1978, Petti 1992, Kohno \etal 2004, Ciangaru \etal 2005).
One reason for the difficulty is that particles in a pencil beam are assumed to receive the same interactions, whereas they may be spatially overreaching beyond the density interface.
The other reason is that only straight paths radiating from a point source are considered in beam transport, whereas actual particles may detour randomly by multiple scattering.

Schneider \etal (1998) showed that a phase-space analysis could address the overreach and detour effects for a simple lateral structure.
Schaffner \etal (1999) and Soukup \etal (2005) subdivided a physical spot beam virtually into smaller beams to naturally reduce overreaches. 
Pflugfelder \etal (2007) quantified lateral heterogeneity, with which subdivision and arrangement could be optimized.
Unfortunately, those techniques are ineffective against beam-size growth during transport.

For electrons, the overreach and detour effects are intrinsically much severer. 
Shiu and Hogstrom (1991) developed a solution, the PB-redefinition algorithm, where minimal pencil beams are occasionally regenerated, considering electron flows rigorously.
The same idea was in fact partly applied to heavy particles for beam customization (Kanematsu \etal 2008b), but the poly-energetic beam model to deal with heterogeneity could be seriously inefficient in high-resolution calculations necessary for Bragg peaks.

In this study, we develop an alternative method to similarly address the overreach and detour effects. 
In the following sections, we incorporate our findings on the Gaussian distribution into the PB algorithm, test the new method in a carbon-ion beam experiment, and discuss the results and practicality for clinical applications.

\section{Materials and methods}

\subsection{Theory}

\subsubsection{Pencil-beam algorithm }\label{sec_pba}

The PB algorithm in this study basically follows our former works (Kanematsu \etal 1998, 2006, 2008b).
A pencil beam with index $b$ is described by position $\vec{r}_b$, direction $\vec{v}_b$, number of particles $n_b$, residual range $R_b$, angular variance $\overline{\theta^2}_b$, angular-spatial covariance $\overline{\theta t}_b$, and spatial variance $\overline{t^2}_b$ of the involved particles.
 As described in \ref{sec_appendix}, these parameters are initialized and modified with transport distance $s$.
The resultant beams with variance $\sigma_b^2 = \overline{t^2}_b$ are superposed to form dose distribution
\begin{eqnarray}
D(\vec{r}) = \sum_{b} \frac{n_b \, D_{\Phi 0}(d_{b r})}{2\, \pi\,\sigma_b^2(s_{b r})}\,\exp\!\left(-\frac{|\vec{r}_{0 b}+s_{b r} \vec{v}_b - \vec{r}|^2}{2\, \sigma_b^2(s_{b r})}\right),
\\
s_{b r} = \left(\vec{r}-\vec{r}_{0 b}\right) \cdot \vec{v}_b,
\qquad
d_{b r} = R_0-R_b(s_{b r}),
\label{eq:convolution}
\end{eqnarray}
where $\vec{r}_{0 b}$ is the beam-$b$ origin, $s_{b r}$ is the distance at the closest approach to point $\vec{r}$, $d_{b r}$ is its equivalent water depth, and $D_{\Phi 0}$ and $R_0$ are the tissue-phantom ratio and the beam range in water.

\subsubsection{Self-similarity of Gaussian distribution }\label{sec_self-similarity}

\begin{figure}
\includegraphics[width=13cm]{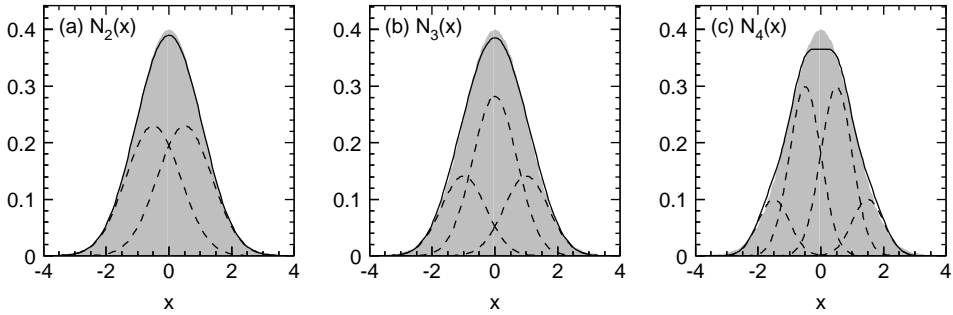}
\caption{The standard normal distribution $N(x)$ (gray area) and its approximate functions (a) $N_2(x)$, (b) $N_3(x)$, and (c) $N_4(x)$ (solid lines) comprised of multiple, displaced, narrowed, and scaled Gaussian distributions (dashed lines).}
\label{fig:splitting}
\end{figure}

Any normalized Gaussian distribution $G_{m,\sigma}(x)$ with mean $m$ and standard deviation $\sigma$ can be represented with the standard normal distribution $N(x) = G_{0,1}(x)$ as
\begin{eqnarray}
G_{m,\sigma}(x) = \frac{1}{\sigma} \, N\!\left(\frac{x-m}{\sigma}\right), 
\qquad
N(x) = \frac{1}{\sqrt{2 \pi}}\, \rme^{-x^2/2}.
\end{eqnarray}
Incidentally, we have found that binomial Gaussian function 
\begin{eqnarray}
N_2(x) = \frac{1}{2}\left[G_{-\frac{1}{2},\frac{\sqrt{3}}{2}}(x) + G_{\frac{1}{2},\frac{\sqrt{3}}{2}}(x)\right],
\end{eqnarray}
reasonably approximates $N(x)$ as shown in \fref{fig:splitting}(a), where we first fixed symmetric displacement $m = \pm 1/2$ for the binomial terms and determined their reduced standard deviation $\sigma = \sqrt{3}/2$ to conserve variance $\int_{-\infty}^{\infty} x^2\, N_2(x)\, \rmd x = 1$.
Similarly, the daughter Gaussian terms in $N_2(x)$ splits into grand daughters to form approximate function
\begin{eqnarray}
N_3(x) = \frac{1}{4} \left[G_{-1,\frac{1}{\sqrt{2}}}(x) + 2\,G_{0,\frac{1}{\sqrt{2}}}(x) + G_{1,\frac{1}{\sqrt{2}}}(x)\right],
\end{eqnarray}
and then into grand-grand daughters to form approximate function
\begin{eqnarray}
N_4(x) = \frac{1}{8}\left[G_{-\frac{3}{2},\frac{1}{2}}(x) + 3\,G_{-\frac{1}{2},\frac{1}{2}}(x) +3\,G_{\frac{1}{2},\frac{1}{2}}(x) +G_{\frac{3}{2},\frac{1}{2}}(x)\right],
\end{eqnarray}
as shown in figures \ref{fig:splitting}(b) and \ref{fig:splitting}(c).
\Tref{tab:splitting} summarizes size-reduction, displacement, and share-fraction factors for splitting with $N_m$ ($m \in \{2, 3, 4\}$).
Further splitting with the same displacement is not possible with valid ($\sigma > 0$) Gaussian terms.

\begin{table}
\caption{Size-reduction, displacement, and share-fraction factors for Gaussian function splitting of multiplicity $m$.}
\begin{indented}
\item[] \begin{tabular}{lcccc}
\br
Factor name & Symbol & $m=2$ & $m=3$ & $m=4$\\
\mr
Size reduction & $\sigma_m$ & $\frac{\sqrt{3}}{2}$ & $\frac{1}{\sqrt{2}}$ & $\frac{1}{2}$\\
Displacement & $\bi{d}_m$ & $\left(\frac{-1}{2},\frac{+1}{2}\right)$ & $\left(-1,0,+1\right)$ & $\left(\frac{-3}{2},\frac{-1}{2},\frac{+1}{2},\frac{+3}{2}\right)$\\
Share fraction & $\bi{f}_m$ & $\left(\frac{1}{2},\frac{1}{2}\right)$ & $\left(\frac{1}{4},\frac{1}{2},\frac{1}{4}\right)$ & $\left(\frac{1}{8},\frac{3}{8},\frac{3}{8},\frac{1}{8}\right)$\\
\br
\end{tabular}
\end{indented}
\label{tab:splitting}
\end{table}

An overreaching Gaussian beam may split two-dimensionally into $m \times m$ smaller beams with these approximations.
Because beam multiplication will explosively increase computational amount, it must be applied only when and where necessary with optimum multiplicity $m$ for required size reduction.

\subsubsection{Lateral heterogeneity}\label{sec_heterogeneity}

\begin{figure}
\indented{\item[]\includegraphics[width=6cm]{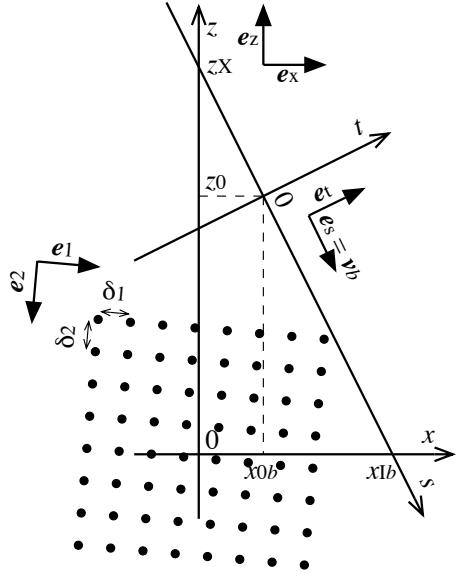}}
\caption{Definitions of the source coordinate system $(x,y,z)$, the beam-$b$ coordinate system $(s,t,u)$, and the density/dose grids at spacing $\delta_1$, $\delta_2$, and $\delta_3$ along axes $\vec{e}_1$, $\vec{e}_2$, and $\vec{e}_3$. The axes normal to the viewing plane, $\vec{e}_u$ and $\vec{e}_3$, are not shown. }
\label{fig:coordinates}
\end{figure}

In a grid-voxel patient model with density distribution $\rho_\mathrm{S}(\vec{r})$, we define density gradient vector $\vec{\nabla}\rho_\mathrm{S}$ as
\begin{eqnarray}
{\vec{\nabla}\rho_\mathrm{S}} = \sum_{g=1}^{3} \frac{\mathrm{maxa}\left[
\rho_\mathrm{S}(\vec{r}+\delta_g\vec{e}_g)-\rho_\mathrm{S}(\vec{r}),\,
\rho_\mathrm{S}(\vec{r})-\rho_\mathrm{S}(\vec{r}-\delta_g\vec{e}_g)\right]}{\delta_g}\, \vec{e}_g,
\end{eqnarray}
where $\delta_g$ and $\vec{e}_g$ are the grid interval and the basis vector for axis $g \in \{1,2,3\}$ as shown in \fref{fig:coordinates} and operation $\mathrm{maxa}[a,b]$ equals $a$ if $|a| \ge |b|$ or otherwise $b$.
We quantify the lateral heterogeneity by effective lateral density gradient
\begin{eqnarray}
\gamma_{xy}(\vec{r}) = \sqrt{\frac{\big| {\vec{\nabla}\rho_\mathrm{S}}\big|^2-\big(\vec{e}_z \cdot {\vec{\nabla}\rho_\mathrm{S}}\big)^2}{2}},
\end{eqnarray}
with which we define the distance to an interface of density change $\kappa_\rho$ as 
\begin{eqnarray}
d_\mathrm{int}(\vec{r}) = \min\!\left( \frac{\kappa_\rho}{\gamma_{xy}(\vec{r})}, 2\,\delta_{xy} \right), 
\end{eqnarray}
where $\kappa_\rho = 0.1$ may be appropriate for interfaces among air ($ \rho_\mathrm{S} \approx 0$), soft tissues ($0.9 \lesssim \rho_\mathrm{S} \lesssim 1.1$), and bones ($1.2 \lesssim \rho_\mathrm{S} \lesssim 1.7$) (Kanematsu \etal 2003).
Effective lateral grid interval
\begin{eqnarray}
\delta_{xy} = \sqrt{\frac{\big|\vec{\delta}_\mathrm{vox}\big|^2 - \big(\vec{e}_z \cdot \vec{\delta}_\mathrm{vox} \big)^2}{2}},
\qquad \vec{\delta}_\mathrm{vox} = \sum_{g=1}^3 \delta_g\,\vec{e}_g,
\end{eqnarray}
multiplied by $2$ is the effective distance to a second laterally adjacent grid, beyond which the distance to the interface can not be estimated from the gradient.

\subsubsection{Beam splitting}\label{sec_splitting}

The pencil beams are examined at every transport step in a patient.
Ones subject to splitting should be not only overreaching beyond a density interface but also substantially influential to the dose for computational efficiency.
We thus require conditions
\begin{eqnarray}
n_b > \kappa_n\, n_{b0},
\qquad R_b > \kappa_R\, R_0, 
\qquad \sigma_b > \frac{\delta_{xy}}{\sqrt{6}},
\qquad \sigma_b > d_\mathrm{int},
\end{eqnarray}
for mother beam $b$ to split.
Cutoff parameters $\kappa_n = 0.1$ and $\kappa_R = 0.1$ set limits on number of particles $n_b$ and residual range $R_b$ with respect to those of the ancestral original beam, $n_{b0}$ and $R_0$.
Condition $\sigma_b > \delta_{xy}/\sqrt{6}$ suppresses splitting into beams narrower than effective grid resolution $\delta_{xy}/\sqrt{12}$ with $m = 2$.
Condition $\sigma_b > d_\mathrm{int}$ defines the state of overreaching.
The optimum multiplicity is determined as
\begin{eqnarray}
m = \cases{
2 & for $\sigma_2\,\sigma_b \leq d_\mathrm{int}$\\
3 & for $\sigma_3\,\sigma_b \leq d_\mathrm{int} < \sigma_2\,\sigma_b$\\
4 & for $\sigma_3\,\sigma_b > d_\mathrm{int}$,\\}
\end{eqnarray}
to suppress recursive splitting with $m = 2$ and 3.
With the beam-$b$ coordinate system $(s,t,u)$ shown in \fref{fig:coordinates} and defined as
\begin{eqnarray}
s = (\vec{r}-\vec{r}_{0 b}) \cdot \vec{e}_s,
\quad t = (\vec{r}-\vec{r}_{0 b}) \cdot \vec{e}_t,
\quad u = (\vec{r}-\vec{r}_{0 b}) \cdot \vec{e}_u,
\\
\vec{e}_s = \vec{v}_b,
\qquad \vec{e}_t =\frac{ v_{b x} \vec{e}_z -v_{b z} \vec{e}_x}{\sqrt{v_{b x}^2+v_{b z}^2}},
\quad \vec{e}_u = \vec{e}_s \times \vec{e}_t,
\end{eqnarray}
daughter beams $b'_{\alpha \beta}$ ($\alpha, \beta \in [1,m]$) are initialized as
\begin{eqnarray}
\vec{r}_{b'_{\alpha \beta}} = \vec{r}_b + \sigma_b \left( {d_m}_\beta\,\vec{e}_t + {d_m}_\alpha\,\vec{e}_u\right),
\\
\vec{v}_{b'_{\alpha \beta}} = \left| \vec{r}_{b'_{\alpha \beta}}-\vec{r}_b + \frac{\overline{t^2}_b}{\overline{\theta t}_b}\,\vec{v}_b \right|^{-1} \left( \vec{r}_{b'_{\alpha \beta}}-\vec{r}_b + \frac{\overline{t^2}_b}{\overline{\theta t}_b}\,\vec{v}_b \right),
\\
n_{b'_{\alpha \beta}} = {f_m}_\alpha\, {f_m}_\beta\, n_b,\qquad
R_{b'_{\alpha \beta}} = R_b,\qquad 
\overline{t^2}_{b'_{\alpha \beta}} = \sigma_m^2\, \overline{t^2}_b,
\\
\overline{\theta t}_{b'_{\alpha \beta}} = \sigma_m^2\, \overline{\theta t}_b, \qquad
\overline{\theta^2}_{b'_{\alpha \beta}} = \overline{\theta^2}_b -\left(1-\sigma_m^2\right) \frac{\overline{\theta t}_b^2}{\overline{t^2}_b},
\end{eqnarray}
where $\vec{r}_{b'_{\alpha \beta}}$ is the displaced position, $\vec{v}_{b'_{\alpha \beta}}$ is the radial direction from the focus or the virtual source (ICRU-35 1984) of the mother beam, $n_{b'_{\alpha \beta}}$ is the number of shared particles, $R_{b'_{\alpha \beta}}$ is the conserved residual range, $\overline{t^2}_{b'_{\alpha \beta}}$ is the reduced spatial variance, and $\overline{\theta t}_{b'_{\alpha \beta}}$ and $\overline{\theta^2}_{b'_{\alpha \beta}}$ conserve focal distance $(\overline{t^2}/\overline{\theta t})$ and local angular variance $(\overline{\theta^2}-\overline{\theta t}^2/\overline{t^2})$ in splitting.
The mother beam splits into the daughter beams to form different detouring paths.

Sets of the initial parameters for daughter beams are sequentially pushed on the stack of computer memory and the last set on the stack will be the first beam to be transported in the same manner, which will be repeated until the stack has been emptied before moving on to the next original beam.

\subsection{Experiment}

\subsubsection{Apparatus }

An experiment to assess the present method was carried out with accelerator facility HIMAC at National Institute of Radiological Sciences.
A $^{12}\mathrm{C}^{6+}$ beam with nucleon kinetic energy $E/A = 290$ MeV was broadened to a uniform field of nominal 10-cm diameter by the spiral-wobbling method (Yonai \etal 2008).
The horizontal wobbler at $z = z_\mathrm{X} = 527$ cm and the vertical wobbler at $z = z_\mathrm{Y} = 470$ cm formed a spiral orbit of maximum 10-cm radius on the isocenter plane.
A 0.8-mm-thick Pb ($\rho_\mathrm{S} = 5.77$, $X_0 = 0.561$ cm) foil was placed at $z = 425$ cm as a scatterer, which increased the instantaneous RMS beam size from pristine 8.3 mm to 25 mm at the isocenter. 
A large-diameter parallel-plate ionization chamber was placed at $z \simeq 400$ cm for dose monitoring and beam-extraction control.
An Al ($\rho_\mathrm{S} = 2.12$, $X_0 = 8.90$ cm) ridge filter for semi-Gaussian range modulation of $m =0.54$ cm and $\sigma = 0.18$ cm in water (Schaffner \etal 2000) and a 2-mm-thick Al base plate were inserted at $z = 235$ cm to moderate the Bragg peak just to ease dosimetry. 

\begin{figure}
\includegraphics[width=13cm]{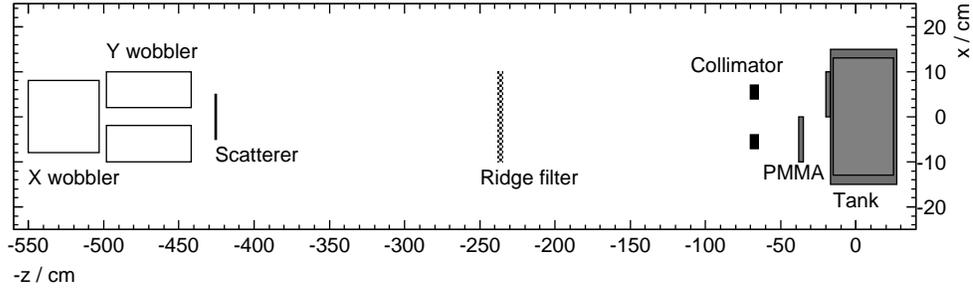}
\caption{Experimental layout of the beam delivery, customization, and phantom systems.}
\label{fig:apparatus}
\end{figure}

As shown in \fref{fig:apparatus}, a water ($\rho_\mathrm{S} = 1$, $X_0 = 36.08$ cm) tank with a 1.9-cm-thick PMMA ($\rho_\mathrm{S} = 1.16$, $X_0 = 34.07$ cm) beam-entrance wall was placed at the irradiation site with the upstream face at $z = 16.9$ cm.
The radiation field was defined by a 8-cm-square 5-cm-thick brass collimator whose downstream face was at 65 cm.
Two identical 3-cm-thick PMMA plates were inserted.
The downstream plate was attached to the beam-entrance face of the tank covering only the $x > 0$ side to form a phantom system with a bump.
The upstream plate was put with its downstream face at $z = 35$ cm covering only the $x < 0$ side to compensate the bump.
Such arrangement is typical for range compensation and sensitive to the detour effects (Kohno \etal 2004).
These beam-customization elements were manually aligned to the nominal central axis at an uncertainty of 1 mm.

A multichannel ionization chamber (MCIC) with 96 vented sense volumes aligned at intervals of 2 mm along the $x$ axis was installed at $y = 0$ in the water tank.
The MCIC system was electromechanically movable along the $z$ axis and the upstream limit at $z_\mathrm{ref} = 14.77$ cm was chosen for the reference point with reference depth $d_\mathrm{ref} = 2.44$ cm of equivalent water from the tank surface.

\subsubsection{Measurement}

With a reference open field without the PMMA plates or the collimator, we measured reference dose/MU reading ${M_\mathrm{ref}}_i$ at reference height $z_\mathrm{ref}$ for every channel $i$ for a calibration purpose.
Every dose/MU reading $M_i(z)$ of channel $i$ at height $z$ for any field is divided by corresponding reference reading ${M_\mathrm{ref}}_i$ to measure dose $D$ at position $(x_i, z)$ as
\begin{equation}
D(x_i,z) = \frac{M_i(z)}{{M_\mathrm{ref}}_ i}\,\frac{z_\mathrm{X}}{z_\mathrm{X}-z_\mathrm{ref}}\,\frac{z_\mathrm{Y}}{z_\mathrm{Y}-z_\mathrm{ref}},
\end{equation}
where divergence-correction factor $z_\mathrm{X}/(z_\mathrm{X}-z_\mathrm{ref})\cdot z_\mathrm{Y}/(z_\mathrm{Y}-z_\mathrm{ref}) = 1.054$ is to measure the doses in dose unit $U$ that would be the isocenter dose for the reference depth of the reference field.

\begin{figure}
\includegraphics[width=13cm]{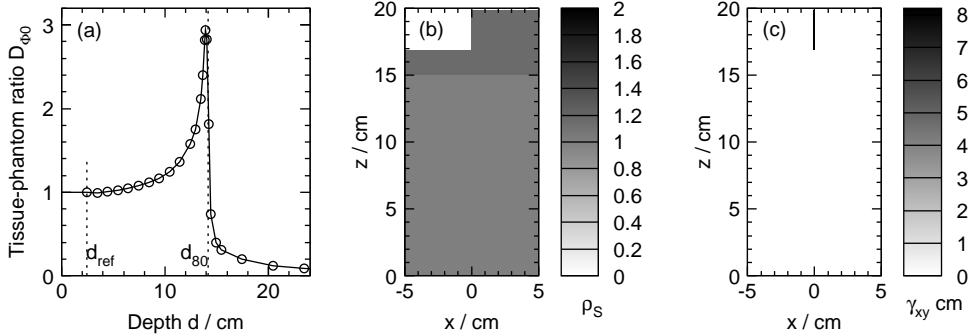}
\caption{(a) Tissue-phantom ratio $D_{\Phi 0}(d)$ with indications for the measurement ($\circ$) and the reference and 80\%-dose depths ($d_\mathrm{ref}$ and $d_{80}$) and (b) effective density $\rho_\mathrm{S}(x,z)$ and (c) effective lateral density gradient $\gamma_{xy}(x,z)$ distributions in gray scale at $y = 0$ in the calculation model.}
\label{fig:model}
\end{figure}

We then measured reference-field doses $D_0(z)$ in the phantom at varied $z$ positions, from which we get tissue-phantom ratio
\begin{equation}
D_{\Phi 0}(d) = D_0(z)\,\frac{z_\mathrm{X}-z}{z_\mathrm{X}}\,\frac{z_\mathrm{Y}-z}{z_\mathrm{Y}},
\qquad
d= d_\mathrm{ref}+z_\mathrm{ref}-z.
\end{equation}
Beam range $R_0$ with Gaussian modulation was equated to the distal 80\%-dose depth (Koehler \etal 1975) $d_{80} = 14.14$ cm as shown in \fref{fig:model}(a).

With the collimator and the PMMA plates in place, lateral dose profiles were measured in the same manner with particular interest around $z =$ 3.3 cm, 6.8 cm, and 10.3 cm, where the Bragg peaks were expected for the primary ions passing through none, either, and both of the PMMA plates.

\subsubsection{Calculation}

\Tref{tab:contributions} shows range loss $-\Delta R$ and scattering $\Delta \overline{\theta^2}$ for the beam-line elements, and the resultant contributions to source sizes $\sigma_\mathrm{X}$ and $\sigma_\mathrm{Y}$ estimated by back projection to the sources.
The ridge filter with the base plate was modeled as plain aluminum of average thickness.
The scattering for the scatterer was estimated from measured beam size 25 mm quadratically subtracted by pristine size in the distance of 425 cm.
Total range loss 2.10 cm was deduced from range 16.24 cm expected for $E/A = 290$ MeV carbon ions (Kanematsu 2008c) and deficit 0.68 cm for the pristine beam may be attributed to minor materials in the beam line.

\begin{table}
\caption{Estimated contributions of beam-line elements at height $z$ to beam range ($R$), scattering angle ($\theta$), and source sizes ($\sigma_\mathrm{X}$ and $\sigma_\mathrm{Y}$).}
\begin{indented}
\item[] \begin{tabular}{lccccc}
\br
Element & $z$ & $-\Delta R$ & $\sqrt{\Delta \overline{\theta^2}}$ & $\sqrt{\Delta \sigma_\mathrm{X}^2}$ & $\sqrt{\Delta \sigma_\mathrm{Y}^2}$ \\
\mr
Pristine & \crule{1} & 0.68 cm & \crule{1} & 8.3 mm & 8.3 mm\\
Scatterer & 425 cm & 0.46 cm & 5.5 mrad & 5.6 mm & 2.5 mm\\
Ridge filter & 235 cm & 0.96 cm & 3.2 mrad & 9.3 mm & 7.5 mm\\
\mr
Total & \crule{1} & 2.10 cm &\crule{1} & 13.7 mm & 11.5 mm\\
\br
\end{tabular}
\end{indented}
\label{tab:contributions}
\end{table}

As described in \ref{sec_appendix}, pencil beams were defined to cover the collimated field at intervals of $\delta_\mathrm{I} = 1$ mm on the isocenter plane, where the open field was assumed to have uniform unit fluence $\Phi_0 = n_b/\delta_\mathrm{I}^2 = 1$.
Exact collimator modeling was omitted because we were interested in the density interface in the middle of the field.
The upstream PMMA plate was modeled as a range compensator with range loss $S = 3.48$ cm for $x < 0$ or $S = 0$ for $x \ge 0$, where the original beams were generated, followed by the range loss and scattering.
The phantom system comprised of the downstream PMMA plate and the water tank was modeled as density voxels at grid intervals of $\delta_1 = \delta_2 = \delta_3 = 1$ mm for a 2-L volume of $|x| \le 5~{\rm cm}$, $|y| \le 5~{\rm cm}$, and $0\le z \le 20~{\rm cm}$.
Figures \ref{fig:model}(b) and \ref{fig:model}(c) show the density and lateral heterogeneity distributions. 

We carried out dose calculations with beam splitting enabled (splitting calculation) and disabled (non-splitting calculation).
In this geometry, the density interface at $x = 0$ was almost parallel to the beams and only ones in the two nearest columns would split. 

\subsection{Applications}

To examine effectiveness and efficiency of this method with larger heterogeneity, a 3-cm diameter cylindrical air cavity at $(x,z) = (-3~{\rm cm}, 13~{\rm cm})$ and two 1-cm diameter bone rods with density $\rho_\mathrm{S} = 2$ at $(2~{\rm cm}, 13~{\rm cm})$ and $(4~{\rm cm}, 13~{\rm cm})$ were added to the phantom in the calculation model.

We carried out splitting and non-splitting dose calculations of the same carbon-ion radiation to monitor changes in frequencies of splitting modes, number of stopped beams, total path length $\sum_b \int \rmd s$, total effective volume $\sum_b \int 12\,\sigma_b^2\, \rmd s$ in the heterogeneous phantom, and computational time with a 2-GHz PowerPC G5/970 processor by Apple/IBM. 

The splitting calculation could be more effective for protons because they generally suffer larger scattering. 
We thus carried out equivalent dose calculations for protons with enhanced scattering angle by factor 3.61 \eref{eq:scattering} in otherwise the same configuration including the tissue-phantom-ratio data.

\section{Results}

\subsection{Experiment}

\begin{figure}
\includegraphics[width=13cm]{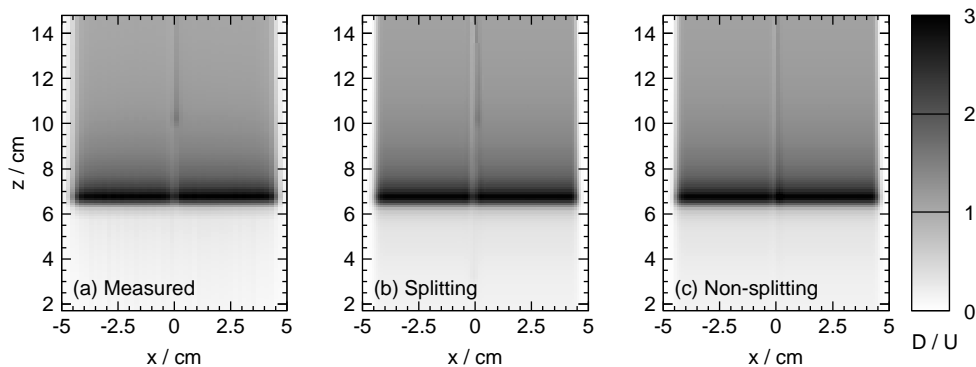}
\caption{Dose distributions in linear gray scale at $y = 0$ by (a) measurement, (b) splitting calculation, and (c) non-splitting calculation.}
\label{fig:distributions}
\end{figure}

\begin{figure}
\includegraphics[width=13cm]{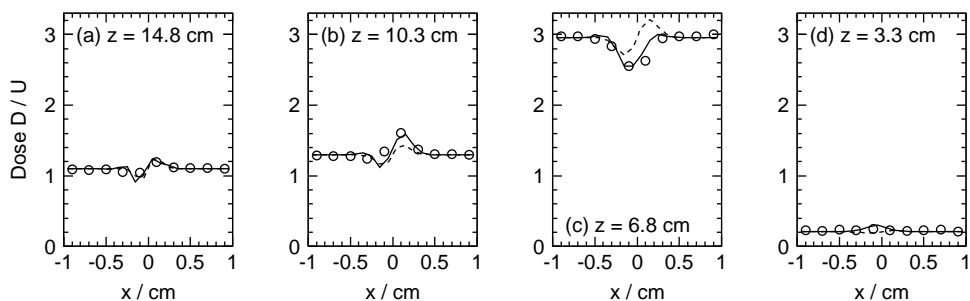}
\caption{Lateral dose profiles at $y = 0$ by measurement ($\circ$), splitting calculation (solid), and non-splitting calculation (dashed) at (a) $z =$ 14.8 cm (plateau), (b) 10.3 cm (sub peak), (c) 6.8 cm (main peak), and (d) 3.3 cm (potential sub peak).} 
\label{fig:profiles}
\end{figure}

\Fref{fig:distributions} shows the two-dimensional dose distributions measured in the carbon-ion beam experiment and the corresponding non-splitting and splitting calculations.
\Fref{fig:profiles} shows their lateral profiles in the plateau and at depths for sub peak, main peak, and potential sub peak expected for particles that penetrated both, either, and none of the PMMA plates.
A dip/bump structure was commonly formed along the $x = 0$ line for lateral particle disequilibrium (Goitein 1978). 
There was actually a sub peak in the measurement and in the splitting calculation, while it was naturally absent in the non-splitting calculation.
The observed loss of the main-peak component was also reproduced by the splitting calculation.
The potential sub peak was barely noticeable only in the splitting calculation.

\subsection{Applications}

\begin{figure}
\includegraphics[width=13cm]{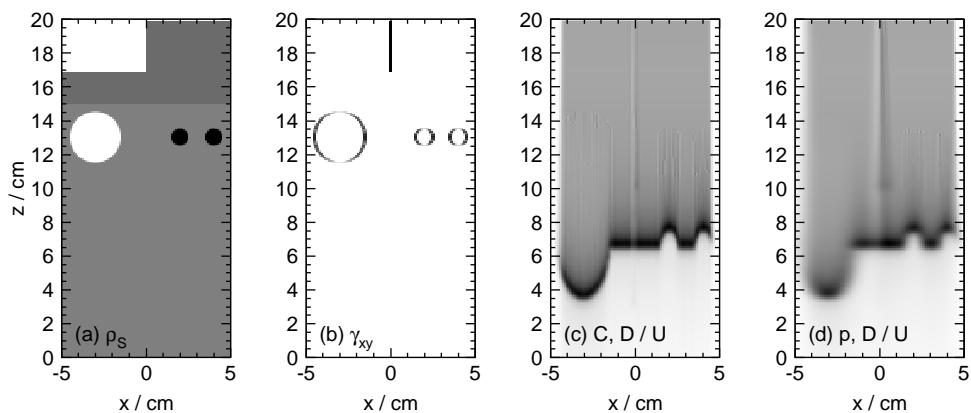}
\caption{Distributions in the heterogeneous phantom at $y = 0$ of (a) density $\rho_\mathrm{S}$ and (b) effective lateral density gradient $\gamma_{xy}$ in the calculation model and doses $D/U$ from (c) carbon-ion and (d) proton radiations calculated with splitting.} 
\label{fig:heterod}
\end{figure}

\begin{figure}
\includegraphics[width=13cm]{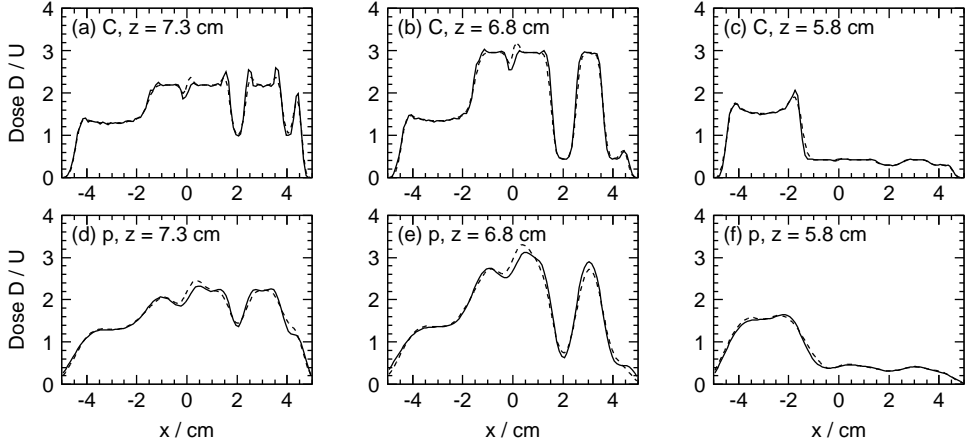}
\caption{Lateral dose profiles in the heterogeneous phantom by splitting (solid) and non-splitting (dashed) calculations for projectile/height of (a) carbon ion/7.3~cm, (b) carbon ion/6.8~cm (main peak), (c) carbon ion/5.8~cm, (d) proton/7.3~cm, (e) proton/6.8~cm (main peak), and (f) proton/5.8~cm.} 
\label{fig:heterop}
\end{figure}

\Fref{fig:heterod} shows details of the heterogeneous phantom and the dose distributions by splitting calculation for the carbon-ion and proton radiations.
The larger scattering for protons naturally led to the larger dose blurring.
\Fref{fig:heterop} shows the dose profiles at the main peak and where the heterogeneity effects were large by splitting and non-splitting calculations.
In addition to the loss of the main-peak component at $x \approx 0$, beam splitting caused some dose enhancement in the shoulders of the profiles especially for the carbon ions.

\Tref{tab:statistics} shows the statistical results, where the splitting effectively increased the carbon-ion and proton beams by factors of 27 and 25 in number, 20 and 25 in path length, 6.6 and 12 in volume, and 4.9 and 4.2 in total computation.

\begin{table}
\caption{Statistics per original beam in splitting and non-splitting calculations for carbon-ion and proton beams in the heterogeneous phantom.}
\begin{indented}\item[]
\begin{tabular}{lcccc}
\br
Projectile & \multicolumn{2}{c}{Carbon ion} & \multicolumn{2}{c}{Proton} \\
Beam splitting & No & Yes & No & Yes \\
\mr
Frequency of $m = 2$ & 0 & 0.243 & 0 & 3.813 \\
Frequency of $m = 3$ & 0 & 0.132 & 0 & 0.714 \\
Frequency of $m = 4$ & 0 & 1.636 & 0 & 0.967 \\
Number of stopped beams & 1 & 26.8 & 1 & 25.0 \\
Mean path length/cm & 20.0 & 394.6 & 20.0 & 499.8 \\
Mean effective volume/cm$^3$ & 3.52 & 23.1 & 30.8 & 380.4 \\
Computational time/s & 9.3 & 45.8 & 15.3 & 63.8 \\
\br
\end{tabular}
\end{indented}
\label{tab:statistics}
\end{table}

\section{Discussion}

Subdivision of a radiation field into virtual pencil beams is an arbitrary process in the PB algorithm although the beam sizes and intervals should be limited by lateral heterogeneity of a given system.
In the PB-redefinition algorithm (Siu and Hogstrom 1991), beams are defined in uniform rectilinear grids and hence regeneration in areas with little heterogeneity may be potentially wasteful.
In the beam-splitting method, beams are automatically optimized in accordance with local heterogeneity. 
In other words, the field will be covered by minimum number of beams in a density-modulated manner as a result of individual independent self-similar splitting.
Relative errors in similarity, $(N_m-N)/N$ ($m \in \{2, 3, 4\}$), are maximum at $x = 0$ with values $-0.9\%$, $-1.3\%$, and $-3.4\%$. 
The resultant dose errors would be smaller, due to contributions of other beams, and may be tolerable. 

Effectiveness of the splitting method was demonstrated in the experiment. The most prominent detour effect was the loss of range-compensated main-peak component in \fref{fig:profiles}(c), which amounted to about 10\% in dose and approximately as large as the distortion due to lateral particle disequilibrium.
The splitting calculation and the measurement generally agreed well, considering that the experimental errors in device alignment could have been 1 mm or more.
The potential sub peak for particles detouring around both PMMA plates was not detected, which may be natural because detouring itself requires scattering.
The dose resolution of the MCIC system of about 1\% of the maximum should have also limited the detectability.

In the applications to the heterogeneous phantom model, although we don't have reference data to compare the results with, it is natural that the splitting calculation with finer beams resulted in finer structures in the dose distributions. 
Computational time is always a concern in practice. 
In our example, the slowing factor for beam splitting with respect to non-splitting calculation was almost common to carbon ions and protons and the speed performance, a minute for 2-L volume in 1-mm grids, may be already acceptable for clinical applications. 

In principle, the total path length determines the computational amount for path integrals \eref{eq:position}--\eref{eq:variance} and the total effective volume determines that for dose convolution \eref{eq:convolution}.
Their Influences on the actual computational time will depend on algorithmic implementations (Kanematsu \etal 2008a). 
In fact, the slowing factor for splitting was less than 5, which is even better than either estimation. 
In addition to common overhead that should have superficially reduced the factor, our code optimization with algorithmic techniques, which will be reported elsewhere, could have contributed to the performance.
Accuracy and speed also depend strongly on the cutoff parameters and logical conditions in the implemented algorithm, size and heterogeneity of a patient model, and resolution clinically needed for a dose distribution.

The automatic multiplication of tracking elements resembles a shower process in physical particle interactions usually calculated in MC simulations.
In fact, MC simulations for dose calculation share many things in common.
Transport and stacking of the elements are essentially the same and the probability for scattering may be equivalent to the distribution in the Gaussian approximation.
As far as efficiency is concerned, the essential differences from the MC method are that the PB method deals with much less number of elements and that it does not rely on stochastic behavior of random numbers.

The beam-splitting method is based on a simple principle of self-similarity and can be applied to any Gaussian beam model of any particle type to fill the gap between Monte Carlo particle simulations and conventional beam calculations in terms of accuracy and efficiency.
However, it is difficult for beam splitting or any beam model in general to deal with interactions that deteriorate particle uniformity, such as nuclear fragmentation processes (Matsufuji \etal 2005).

\section{Conclusions}

In this work, we applied our finding of self-similar nature of Gaussian distributions to dose calculation of heavy charged particle radiotherapy.
The self-similarity enables dynamic, individual, and independent splitting of Gaussian beams that have been grown larger than the limit from lateral heterogeneity of the medium. 
As a result, pencil beams will be arranged with optimum and modulated areal density to minimize overreaching and to address detouring with deflecting daughter beams. 

In comparison with a conventional calculation and a measurement, the splitting calculation was prominently effective in the target region with steep range adjustment by an upstream range compensator. 
The detour effect was about 10\% for the maximum and of the same order of magnitude with lateral particle disequilibrium effect.
In comparison between carbon ions and protons, the effects of splitting were not significantly different because other scattering effects were also larger for protons.

Although performances depend strongly on physical beam conditions, clinical requirement, and algorithmic implementation, a typical slowing factor of the order of 10 may be reasonably achievable for involvement of beam splitting. 
In fact, factor of 5 has been achieved in our example. 
The principle and formulation for beam splitting are general and thus the feature may be added to various implementations of the PB algorithm in a straightforward manner.

\appendix
\section{Pencil beam generation and transport}\label{sec_appendix}

On generation of pencil beam $b$ on a plane at height $z_0$ as shown in \fref{fig:coordinates}, beam position $\vec{r}_b$, residual range $R_b$, and variances $\overline{\theta^2}_b$, $\overline{\theta t}_b$, and $\overline{t^2}_b$ are initialized as
\begin{eqnarray}
\vec{r}_b(0) = \vec{r}_{0 b} = \vec{r}_{\mathrm{I} b}+\frac{z_0}{ v_{b z}} \vec{v}_b,
\qquad
R_b(0) = R_0,
\\
\overline{\theta^2}_b(0) = \frac{1}{2} \left(\frac{\sigma_\mathrm{X}}{z_\mathrm{X}-z_0}\right)^2+ \frac{1}{2} \left(\frac{\sigma_\mathrm{Y}}{z_\mathrm{Y}-z_0}\right)^2, 
\\
\overline{t^2}_b(0) = \frac{z_\mathrm{X}-z_0 }{ z_\mathrm{X}}\,\frac{z_\mathrm{Y}-z_0 }{ z_\mathrm{Y}}\,\frac{\delta_\mathrm{I}^2}{12}, 
\qquad 
\overline{\theta t}_b(0) = \frac{\overline{t^2}_b}{\sqrt{z_\mathrm{X}-z_0}\sqrt{z_\mathrm{Y}-z_0}},
\end{eqnarray}
where $\vec{r}_{0 b} = (x_{0 b}, y_{0 b}, z_0)$ is the beam-$b$ origin, $\vec{r}_{\mathrm{I} b} = (x_{\mathrm{I} b}, y_{\mathrm{I} b}, 0)$ is the beam position on the isocenter plane, $\sigma_\mathrm{X}$ and $\sigma_\mathrm{Y}$ are the source sizes at virtual source heights $z_\mathrm{X}$ and $z_\mathrm{Y}$, $R_0$ is the initial residual range, and $\vec{v}_b = (v_{b x}, v_{b y}, v_{b z})$ is the beam direction radiating from the virtual sources with
\begin{eqnarray}
\frac{ v_{b x}}{ v_{b z}} = -\frac{x_{\mathrm{I} b}}{z_\mathrm{X}},
\quad \frac{ v_{b y}}{ v_{b z}} = -\frac{y_{\mathrm{I} b}}{z_\mathrm{Y}},
\quad v_{b z} = -\left(\frac{x_{\mathrm{I} b}^2}{z_\mathrm{X}^2}+\frac{y_{\mathrm{I} b}^2}{z_\mathrm{Y}^2}+1\right)^{-\frac{1}{2}}.
\end{eqnarray}
Because nuclear interactions are effectively handled in tissue-phantom ratio $D_{\Phi 0}(d)$ in dose calculation, number of particles $n_b$ is modeled as invariant.

The Fermi-Eyges theory (Eyges 1948, Kanematsu 2008c, 2009) gives increments of the PB parameters in step $\Delta s$ within a density voxel by
\begin{eqnarray}
\Delta \vec{r}_b = \vec{v}_b\, \Delta s, \label{eq:position}
\qquad \Delta R_b = -\rho_\mathrm{S}\, \Delta s, \label{eq:range}
\\
\Delta \overline{\theta^2}_b = \left(1.00 \times 10^{-3}\right) z^{-0.16} \left(\frac{m}{m_p}\right)^{-0.92} \frac{X_{0 \mathrm{w}}}{\rho_\mathrm{S}\,X_0} \ln\frac{R_b}{R_b+\Delta R_b}, \label{eq:scattering}
\\
\Delta \overline{\theta t}_b = \left( \overline{\theta^2}_b + \frac{\Delta \overline{\theta^2}_b}{2}\right) \Delta s,
\\
\Delta \overline{t^2}_b = \left[ 2\, \overline{\theta t}_b + \left( \overline{\theta^2}_b + \frac{\Delta \overline{\theta^2}_b}{3}\right) \Delta s \right] \Delta s, \label{eq:variance}
\end{eqnarray}
where $\rho_\mathrm{S}$ and $X_0/X_{0 \mathrm{w}}$ are the effective density (Kanematsu \etal 2003) and radiation length of the medium in units of those of water and $z$ and $m/m_p$ are the particle charge and mass in units of those of a proton.
For the last physical step with $R_b \to 0$ and diverging $\Delta \overline{\theta^2}_b$, the growth is directly given by $\Delta \overline{t^2}_b = 0.0224^2 z^{-0.16} (m/m_p)^{-0.92} (R_b/\rho)^2$ and then disabled by $\Delta \overline{t^2}_b = 0$ in the unphysical $R_b \leq 0$ region.

\References

\item[] Ciangaru G, Polf J C, Bues M and Smith A R 2005 Benchmarking analytical calculations of proton doses in heterogeneous matter {\it Med. Phys.} {\bf 32} 3511--23

\item[] Eyges L 1948 Multiple scattering with energy loss {\it Phys. Rev.} {\bf 74} 1534--5

\item[] Goitein M 1978 A technique for calculating the influence of thin inhomogeneities on charged particle beams {\it Med. Phys.} {\bf 5} 256--264

\item[] Hong L, Goitein M, Bucciolini M, Comiskey R, Gottschalk B, Rosenthal S, Serago C and Urie M 1996 A pencil beam algorithm for proton dose calculations \PMB {\bf 41} 1305--30

\item[] ICRU-35 1984 Radiation dosimetry: electron beams with energies between 1 and 50 MeV {\it ICRU Report} 35 (Bethesda, MD: ICRU)

\item[] Kanematsu N, Akagi T, Futami Y, Higashi A, Kanai T, Matsufuji N, Tomura H and Yamashita H 1998 A proton dose calculation code for treatment planning based on the pencil beam algorithm {\it Jpn. J. Med. Phys.} {\bf 18} 88--103

\item[] Kanematsu N, Matsufuji N, Kohno R, Minohara S and Kanai T 2003 A CT calibration method based on the polybinary tissue model for radiotherapy treatment planning \PMB {\bf 48} 1053--64

\item[] Kanematsu N, Akagi T, Takatani Y, Yonai S, Sakamoto H and Yamashita H 2006 Extended collimator model for pencil-beam dose calculation in proton radiotherapy \PMB {\bf 51} 4807--17

\item[] Kanematsu N, Yonai S and Ishizaki A 2008a The grid-dose-spreading algorithm for dose distribution calculation in heavy charged particle radiotherapy {\it Med. Phys} {\bf 35} 602--7

\item[] Kanematsu N, Yonai S, Ishizaki A and Torikoshi M 2008b Computational modeling of beam-customization devices for heavy-charged-particle radiotherapy \PMB {\bf 53} 3113--27

\item[] Kanematsu N 2008c Alternative scattering power for Gaussian beam model of heavy charged particles {\it Nucl. Instrum. Methods} B {\bf 266} 5056--62

\item[] Kanematsu N 2009 Semi-empirical formulation of multiple scattering for Gaussian beam model of heavy charged particles stopping in tissue-like matter \PMB (under review)

\item[] Koehler A M, Schneider R J and Sisterson J M 1975 Range modulators for protons and heavy ions {\it Nucl. Instrum. Methods} {\bf 131} 437--40

\item[] Kohno R, Kanematsu N, Kanai T and Yusa K 2004 Evaluation of a pencil beam algorithm for therapeutic carbon ion beam in presence of bolus {\it Med. Phys.} {\bf 31} 2249--53

\item[] Kr\"amer M, J\"akel O, Haberer T, Kraft G, Schardt D and Weber U 2000 Treatment planning for heavy-ion radiotherapy: physical beam model and dose optimization \PMB {\bf 45} 3299--317

\item[] Matsufuji N, \etal 2005 Spatial fragment distribution from a therapeutic pencil-like carbon beam in water \PMB {\bf 50} 3393--403

\item[] Petti P L 1992 Differential-pencil-beam dose calculations for charged particles {\it Med. Phys.} {\bf 19} 137--49

\item[] Pflugfelder D, Wilkens J J, Szymanowski H and Oelfke U 2007 Quantifying lateral heterogeneities in hadron therapy {\it Med. Phys.} {\bf 34} 1506--13

\item[] Schaffner B, Pedroni E and Lomax A 1999 Dose calculation models for proton treatment planning using a dynamic beam delivery system: an attempt to include density heterogeneity effects in the analytical dose calculation \PMB {\bf 44} 27--41

\item[] Schaffner B, Kanai T, Futami Y, Shimbo M and Urakabe E 2000 Ridge filter design and optimization for the broad-beam three-dimensional irradiation system for heavy-ion radiotherapy {\it Med. Phys.} {\bf 27} 716--24

\item[] Schneider U, Schaffner B, Lomax T, Pedroni E and Tourovsky A 1998 A technique for calculating range spectra of charged particle beams distal to thick inhomegeneities {\it Med. Phys.} {\bf 25} 457--63

\item[] Shiu A S and Hogstrom K R 1991 Pencil-beam redefinition algorithm for electron dose distributions {\it Med. Phys.} {\bf 18} 7--18

\item[] Soukup M, Fippel M and Alber M 2005 A pencil beam algorithm for intensity modulated proton thepray derived from Monte Carlo simulations \PMB {\bf 50} 5089--104

\item[] Yonai S, Kanematsu N, Komori M, Kanai T, Takei Y, Takahashi O, Isobe Y, Tashiro M, Koikegami H and Tomita H 2008 Beam wobbling methods for heavy-ion radiotherapy {\it Med. Phys.} {\bf 35} 927--38.

\endrefs

\end{document}